\documentclass[prb,twocolumn,amsmath,floats,showpacs]{revtex4}

\usepackage{graphicx}

\begin{document}
\input epsf

\title{Point Contact Spectroscopy of Superconducting Gap Anisotropy in Nickel
Borocarbide Compound LuNi$_{2}$B$_{2}$C}

\author{N.~L.~Bobrov $^{1,2}$,
S.~I.~Beloborod'ko $^{1}$, L.~V.~Tyutrina $^{1}$, and I.~K.~Yanson
$^{1}$\footnote{Email address: yanson@ilt.kharkov.ua} }
\affiliation{$^{1}$B.I.~Verkin Institute for Low Temperature
Physics and Engineering, NAS of Ukraine, 47, Lenin Prospect,
61103, Kharkiv, Ukraine}

\author {D.~G.~Naugle $^{2}$ and K.~D.~D.~Rathnayaka $^{2}$}
\affiliation{$^{2}$Department of Physics Texas A\&M University,
College Station TX 77840-4242, USA}

\date{\today}

\begin{abstract}
Point contacts are used to investigate the anisotropy of the
superconducting energy gap in LuNi$_{2}$B$_{2}$C in the $ab$ plane
and along the $c$ axis. It is shown that the experimental curves
should be described assuming that the superconducting gap is
non-uniformly distributed over the Fermi surface. The largest and
the smallest gaps have been estimated by two-gap fitting models.
It is found that the largest contribution to the point-contact
conductivity in the $c$ direction is made by a smaller gap and,
in the $ab$ plane by a larger gap. The deviation from the one-gap
BCS model is pronounced in the temperature dependence of the gap
in both directions. The temperature range, where the deviation
occurs, is for the $c$ direction approximately 1.5 times more than
in the $ab$ plane. The $\Gamma$ parameter, allowing quantitatively
estimate the gap anisotropy by one-gap fitting, in $c$ direction
is also about 1.5 times greater than in the $ab$ plane. Since it
is impossible to describe satisfactorily such gap distribution
either by the one- or two-gap models, a continuous, dual-maxima
model of gap distribution over the Fermi surface should be used
to describe superconductivity in this material.

\pacs{63.20.Kr, 72.10.Di, 73.40.Jn}
\end{abstract}

\maketitle

\section{Introduction}

The goal of this study was to investigate the anisotropy of the
energy gap in nickel borocarbide superconductors {\it
Re}Ni$_{2}$B$_{2}$C.  The crystallographic structure of these
compounds resembles to some extent the structure of HTSC
materials.\cite{Volkova} It consists of alternating {\it Re}-C and
Ni$_2$-B$_2$ layers ({\it Re} is a rare earth metal). There is a
point of view that the electron properties of {\it
Re}Ni$_{2}$B$_{2}$C compounds in the normal state are isotropic
because of the strong carbon bond along the tetragonal $c$
axis.\cite{Muller} This is supported by the temperature dependence
of isotropical resistivity in the single crystal of
YNi$_{2}$B$_{2}$C.\cite{Daya} On the other hand, there is also
data about a substantial anisotropy in these
compounds.\cite{K.Muller} In the superconducting state the
experimental data are also contradictory. The upper critical
fields for $H$ parallel and perpendicular to the $c$ axis and the
derived superconducting parameters do not show any anisotropy for
the YNi$_{2}$B$_{2}$C single-crystal samples in agreement with
magnetization and torque magnetometry measurements, but a small
anisotropy is observed for the LuNi$_{2}$B$_{2}$C single
crystals.\cite{Daya,Metlushko} Meanwhile, a very strong anisotropy
of the superconducting energy gap was reported for
LuNi$_{2}$B$_{2}$C in the $ab$ plane.
This conclusion was based on the observation of delocalized
quasiparticles in thermal conductivity at very low
temperatures.\cite{Boaknin} According to this analysis, the
smallest gap differs at least ten times from the gap in other
directions. Recently, evidence for the presence of nodes along
$\langle100\rangle$ direction was provided by the field-angle
thermal conductivity,\cite{Izawa} field-angle heat
capacity\cite{Park} and ultrasonic attenuation measurements of
YNi$_{2}$B$_{2}$C.\cite{Watanabe}

In accordance with the experimental data an anisotropic
superconducting gap function was proposed in
Ref.\,\onlinecite{Maki} (see, also, Fig.\,52 from the recent
review\cite{review}). In that model the gap has nodes along
$\langle100\rangle$ directions and attains the maximum values
along $\langle110\rangle$ directions. In the case of $s+g$
symmetry, the elastic scattering leads both to a decrease of $T_c$
and to an "isotropization" of the gap with vanishing nodes.

We do not know any direct measurement of gap anisotropy involving
the Andreev reflection.\cite{Blonder} The STM tunnelling
measurements at 4\,K report a gap of 2.2 meV along the $c$ axis,
yielding a too low ratio $2\Delta /kT_c=3.2$ for
LuNi$_{2}$B$_{2}$C.\cite{DeWilde1,DeWilde2} By contrast, there are
point-contact measurements of the gap in the $ab$ plane ($e.g.$
\cite{Bobrov,Rybalchenko}), which yield $2\Delta
/kT_c=3.7\div3.8$. The point-contact spectroscopy method in the
latter measurements cannot, however, provide angular resolution
of the gap anisotropy much better than $\pi/2$.

In the present investigation we have found that both in the $ab$
plane and in $c$ direction the experimental point-contact spectra
cannot be fitted satisfactory with the one-gap theoretical curve,
even when broadened by an adjustable broadening parameter.
Fitting the experimental spectra both at a small bias
($eV<\Delta$) and at a bias larger than $\Delta$ forces us to use
at least the two-gap fitting curve like the one for the two-band
superconductor.\cite{Blonder,Brinkman} But even the two-gap
fitting with the two proper broadening parameters was not good
enough to approximate the experimental characteristic at the
middle energy region ($eV\sim\Delta$). Moreover, for the two-band
model one could expect transitions of Cooper pairs between the
bands to introduce an additional depairing factor, which is
analogous to other depairing factors like magnetic impurities,
strong magnetic fields, etc. In such case the superconducting
order parameter and the energy gap should differ from each other,
which could be accounted for by using
Ref.\,\onlinecite{Beloborodko}. Unfortunately even using the
latter theory satisfactory fits could not be obtained. Therefore
we suggest that only a continuous gap distribution  with the two
energy gap maxima around the Fermi surface could satisfactorily
approximate the experimental spectra at low temperature
($T<<T_c$). We have determined tentatively the value of these
maxima and the range of gap distribution which corresponds well
to the recent STM measurements.\cite{Martinez2}

\section{Experimental technique}

The point contact measurement was performed on single crystal
LuNi$_{2}$B$_{2}$C with $T_{c}\simeq16.9$ K grown by Canfield and
Bud'ko using a flux method.\cite{Cho} Geometrically, our crystal
was a thin ($0.1 \sim 0.2$ mm) plate with the $c$ axis
perpendicular to its plane. The single crystal surface always
contains quite a thick layer in which superconductivity is either
absent or strongly suppressed. To perform measurement in the $ab$
plane, the crystal is usually cleaved and the point contact is
made between a metallic counterelectrode and the cleaved surface.
It was technically problematic to produce a cleavage perpendicular
to the $c$ direction. In this case the crystal surface was cleaned
with a 10\% HNO$_{3}$ solution in ethanol. As the measurement in
the $ab$ plane shows, both cleavage and etching yield identical
results.

The point contact in the  $ab$ plane is fabricated between an edge
of the silver electrode and a freshly cleaved (or etched)
corresponding facet of the single crystal.\cite{Chubov} The
deviation from the perpendicular to the $c$ direction might amount
about $5\div10^\circ$. We do not know {\it a priori} along which
of the in-plane directions the contact is obtained. But since we
used a selection rule to choose the highest observable
superconducting energy gap with the largest nonlinearity at the
gap double-minimum structure, the contact axis is presumably along
$\langle110\rangle$ directions, where the in-plane gap is maximum.
To produce contacts in the $c$ direction the traditional
"needle-anvil" geometry is used. The radius of the needle is about
1-3 micron. The temperature was measured using a special cryogenic
insert (its close analogue is described elsewhere.\cite{Engel})

The point contact resistances varied typically from several Ohms
to tens of Ohms. For detailed investigations we chose the point
contacts with the highest permissible tunneling which was
controlled against the differential resistance maximum at zero
bias and by the maximum nonlinearity in the dV/dI-double-minimum
region, corresponding to the intact superconducting surface under
the contact. Unfortunately, a complete set of curves in the whole
$T_{min}-T_{c}$ interval was obtained only for a few contacts.
Because of their rather high resistance and the long-duration
(over 10-12 hours) of the measurements, many of the contacts were
broken down. The temperature interval ranges from the minimal
available $T$ ($\sim$1.5 K) to $T_c$.

\section{Results and discussion}

\begin{figure}[htb]
\includegraphics[width=8cm,angle=0]{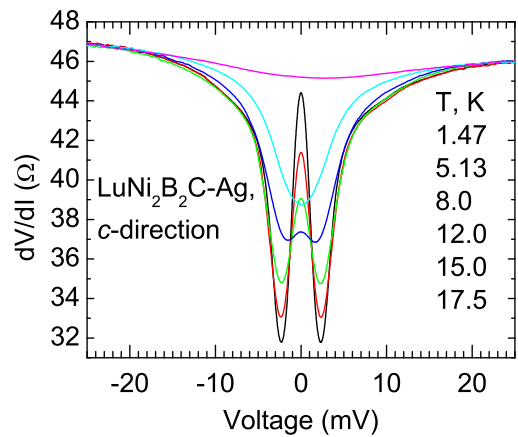}
\caption[]{ Differential resistance of LuNi$_{2}$B$_{2}$C-Ag point
contact at different temperatures. The contact axis is parallel to
$c$. Not to overload the figure, we give only several
representative curves.} \label{fig1}
\end{figure}

\begin{figure}[htb]
\includegraphics[width=8cm,angle=0]{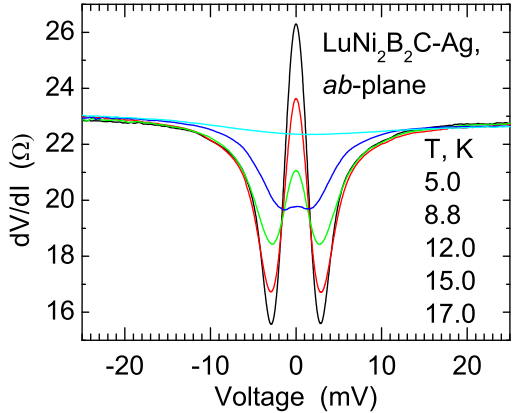}
\caption[]{ Differential resistance of LuNi$_{2}$B$_{2}$C-Ag point
contact at different temperatures. The contact axis is along the
$ab$ plane. See the last sentence in the previous caption.}
\label{fig2}
\end{figure}

Some curves of the first derivatives $dV/dI$ measured  at
different temperatures on LuNi$_{2}$B$_{2}$C-Ag point contacts
along $c$ axis and in the $ab$ plane are shown in
Figs.\,\ref{fig1},\ref{fig2}.


The measured curves were symmetrized
$dV/dI_{sym}=1/2\,[dV/dI(V)+dV/dI(-V)]$ and normalized to the
normal state at $T>T_{c}$, except in Figs\,1,2, where the raw
data are shown. Each curve contains several hundred of
experimental dots and at the scale shown in presented figures,
all the curves have negligible noise. In other figures
experimental curves are shown by dots whose number is decreased
in order to be discerned. Statistically (several tens of contacts
were examined for each direction), the distance between the
minima in the first derivative dV/dI characterizing the average
value  of the gap was 12-15\% larger in the $ab$ plane compared
to that in the $c$ direction. This is a very crude estimation of
the $c\leftrightarrow ab$ anisotropy.

The theoretical predictions and the experimental results were
compared using two approaches. First, a model was applied, which
describes electrical conductivity of pure S-c-N point contacts in
the presence of an arbitrarily transparent potential barrier at
the boundary between the metals. This model allows for the finite
lifetime of the Cooper pairs.\cite{Beloborodko} The $I-V$
characteristics of the point contact are described as follows:

\begin{eqnarray}
I = \frac{{1}}{{2eR_{N}} }\int\limits_{0}^{\infty} d\varepsilon
\left[ {\tanh\frac{{\varepsilon + eV}}{{2T}} -
\tanh\frac{{\varepsilon - eV}}{{2T}}} \right] \nonumber \\
\times \frac{{1}}{{\left| {\left( {2d^{ - 1} - 1} \right) +
g_{\varepsilon} ^{R}} \right|^{2}}} \label{eq1} \\
\times\left\{ {\left| {1 + g_{\varepsilon }^{R}} \right|^{2} +
\left| {f_{\varepsilon} ^{R}} \right|^{2} + 4\left( {d^{ - 1} - 1}
\right)\text{Re}g_{\varepsilon} ^{R}}  \right\},\nonumber
\end{eqnarray}
where $g_{\varepsilon} ^{R} = \frac{u}{ \sqrt{u^{2} - 1}}$; \quad
$f_{\varepsilon} ^{R} = \frac{1}{\sqrt{u^{2} - 1}}$;
$d=\frac{{1}}{1 + Z^{2}}$, $d$ is the transparency of the
potential barrier, and $Z$ is the tunneling parameter. The value
$u=\varepsilon/\Delta_{\varepsilon}$ ($\Delta_{\varepsilon}$ - is
a complex energy-independent order parameter of the
superconductor), which can be found from the
equation:\cite{Abrikosov}

\begin{equation}
u = \frac{{\varepsilon} }{{\Delta} } + i\gamma
\frac{u}{\sqrt{u^{2} - 1}}, \label{eq2}
\end{equation}

\noindent where $\Delta$ is the BCS order parameter obtained from
\begin{equation}
 \Delta = \lambda \int\limits_{0}^{\omega _{D}}
{d\varepsilon \tanh\left( {\frac{{\varepsilon} }{{2T}}}
\right)\text{Re}\frac{1}{{\sqrt {u^{2}-1}} }} ,
\end{equation}

\noindent and $\gamma=1/\tau_{S}\Delta$ is the pair breaking
parameter ($\tau_{S}$ is the mean free time during, $e.g$.,
spin-flip scattering at impurities). When magnetic impurities are
absent, $\tau_{S}$ tends to infinity and Eq.\,(1) coincides with
the equation for current in Ref.\,\onlinecite{Zaitsev}. The
energy gap $\Delta_{0}$ and the order parameter $\Delta$ are
related as follows:

\begin{equation}
\label{eq3} \Delta _{0} = \Delta \left( {1 - \gamma ^{2/3}}
\right)^{3/2}.
\end{equation}

The other approach was based on the generalized
Blonder-Tinkham-Klapwijk (BTK) model\cite{Plecenik} commonly used
to describe S-c-N point contacts. The model allows for the finite
lifetime of quasiparticles $\tau$=$\hbar/\Gamma$ determined by
inelastic scattering, which leads to the broadening of the density
of states in the superconductor. Formally, according to the theory
\cite{Beloborodko} the BTK-based results are obtained under the
condition of strong pair breaking ($|{u}|$$\gg$1). Therefore, in
the strict sense, the generalized BTK model contains no gap. For
any infinitesimal broadening parameter $\Gamma$, at $T=0$ the
density of states near the Fermi surface is non-zero. In
theory,\cite{Beloborodko} the order parameter is a quantity
analogous to the pseudogap in the generalized BTK model. In the
following we use terms "order parameter" and $\gamma$ for the
model of Ref.\onlinecite{Beloborodko}, while the "gap" and
$\Gamma$ for the commonly used BTK model,\cite{Plecenik} except
special cases, where we refer to $\Delta _{0}$ related to the
model.\cite{Beloborodko}

Now we consider more closely the methods of fitting the
theoretical curves to the experimental results. The iteration method
commonly used in this case is quite good for small broadening.
However, when the broadening $\Gamma$ is comparable with the gap
(or $\gamma>$ 0.3), the results thus obtained are ambiguous and
often dependent on the starting $\Delta, \Gamma$(or $\gamma$) and
$Z$ values. In our calculation, we therefore used the technique of
coordinate descent with a postponed solution.

First, we specified an interval in which $\Delta$ is searched at a
given temperature. The interval was then subdivided into
equidistant parts: $\Delta_{1}, \Delta_{2}, ... …\Delta_{n}$.
$\gamma$ and $Z$ were fitted for each $\Delta_{i}$. The procedure
was as follows: after each step of $\gamma$ -fitting, a complete
$Z$-fitting was performed and then the next step of
$\gamma$-fitting was considered. Before each calculation of the
average r.m.s. deviation F$(\Delta _i)$, the amplitudes of the
fitting and the experimental curves were made equal by multiplying
the $y$-coordinate of the fitting curve by a scale-factor $S$ and
then shifting it along the $y$-axis by an amount $B$. The values
of $B$ and $S$ were chosen to minimize the deviation F($\Delta
_i$). The standard algorithm for determination of $B$ and $S$ is
considered, for instance, in Ref.\,\onlinecite{algorithm}. As a
result, for each $\Delta_{i}$ we found $\gamma_{i}$ and $Z_{i}$ at
which the difference between the shapes of the theoretical and
experimental curves characterized by the r.m.s. deviation
F$(\Delta_{i}$) was the smallest one.
\begin{figure}[htb]
\includegraphics[width=8cm,angle=0]{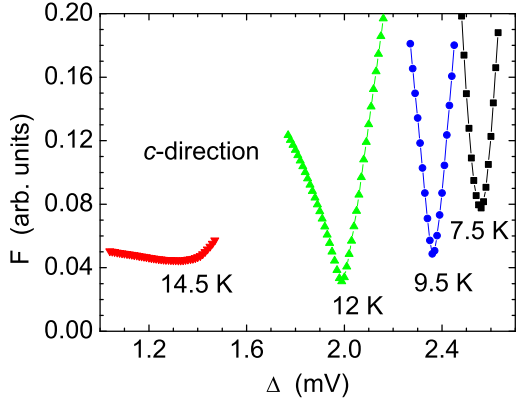}
\caption[]{Averaged r.m.s. deviation F characterizing the extent
of shape discrepancy between the theoretical and experimental
curve as a function of $\Delta$. The minima in the curves
correspond to the best agreement of theory and experiment. The
data shown are for LuNi$_{2}$B$_{2}$C-Ag point contacts in the $c$
direction at different temperatures. Calculation based on
Ref.\,\onlinecite{Beloborodko}.} \label{fig3}
\end{figure}
\begin{figure}[htb]
\includegraphics[width=8cm,angle=0]{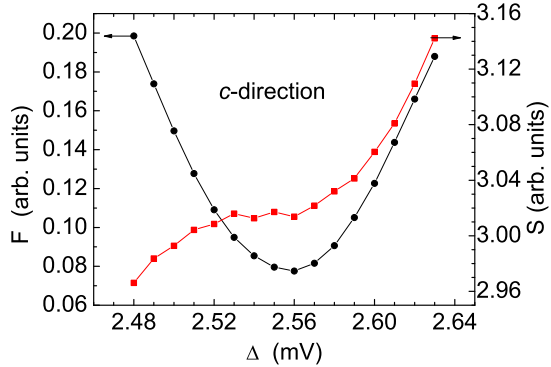}
\caption[]{$\Delta$-dependences of the averaged r.m.s. deviation
{\it F} (see Fig.\,\ref{fig3}) and the scale factor {\it S} at
$T=7.5$~K. Calculation by equations from
Ref.\,\onlinecite{Beloborodko}. $S$ is a factor used to divide the
$y$-coordinate of the theoretical curve to match its amplitude
with that of the experimental curve.} \label{fig4}
\end{figure}
\noindent The same holds for $\Gamma$ in the BTK fitting
model.\cite{Plecenik} The calculation for some temperatures is
shown in Fig.\,\ref{fig3}. It is evident that at $T_{c}$ the
technique cannot ensure unambiguous results. At 14.5~K, the curve
F$(\Delta_{i}$) is practically horizontal. The determination of
the $\Delta$-value should take into account other factors as well.
For example, it is important that the tunneling parameter $Z$ and
the scale factor $S$ should be invariant. The dependences
S$(\Delta$) and F$(\Delta$) at $T=7.5$~K are shown in
Fig.\,\ref{fig4}. It is seen that the minimum F corresponds to a
certain S-value. The BTK calculation of the corresponding
dependences at the same temperature yields broader F$(\Delta$)
curves. Nevertheless, the minima positions in these curves are
quite definite.

\begin{figure}[htb]
\includegraphics[width=8cm,angle=0]{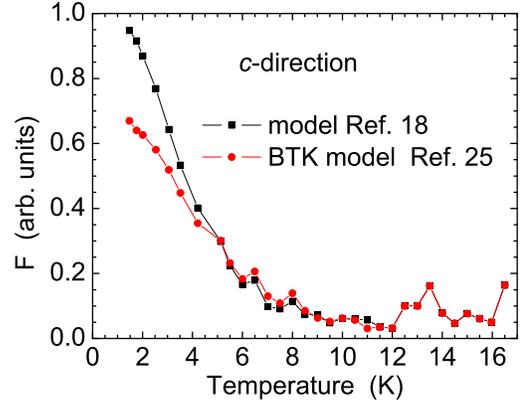}
\caption[]{Temperature dependence of the shape discrepancy between
theoretical and experimental curves calculated according to
Ref.\,\onlinecite{Beloborodko} and within the generalized BTK
model \cite{Plecenik} for LuNi$_{2}$B$_{2}$C-Ag point contacts in
the $c$ direction.} \label{fig5}
\end{figure}

At low temperatures the shapes of the theoretical and the
experimental curves were in rather poor agreement for all of the
borocarbides investigated ({\it Re}=Er, Dy, Tm, Lu) irrespective
of the point contact orientation. The temperature dependence of
the r.m.s. deviation in shape in the $c$ direction is shown in
Fig.\,\ref{fig5} for both models.\cite{Beloborodko,Plecenik}
\begin{figure}[htb]
\includegraphics[width=8cm,angle=0]{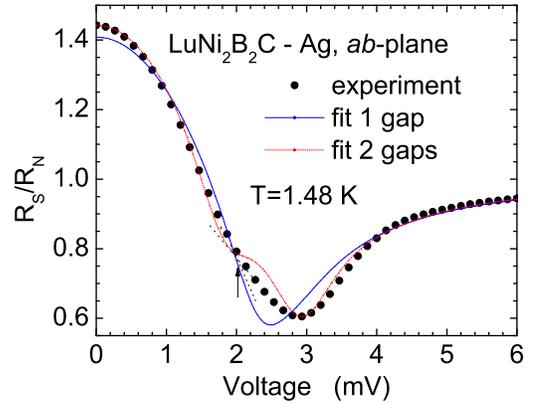}
\caption[]{Approximation of the experimental curve with one- and
two-gap models for the LuNi$_2$B$_2$C-Ag point contact in the $ab$
plane. The dotted straight lines marked with an arrow indicate the
kink in the experimental curve which is exaggerated by the 2-gap
theoretical fit. } \label{fig6}
\end{figure}
It is seen that the best agreement is achieved above 7 K.  At
$T<$7 K, the largest discrepancy between the shapes is observed in
the region of extrema. The central maximum of the differential
resistance in the experimental curve is considerably narrower than
that in the best fit,\footnote{"fit" stands for the theoretical
$dV/dI(V)$-dependence fitted to the experimental results.}
whereas the minima are shifted to higher bias (Fig.\,\ref{fig6}).

\begin{figure}[htb]
\includegraphics[width=8cm,angle=0]{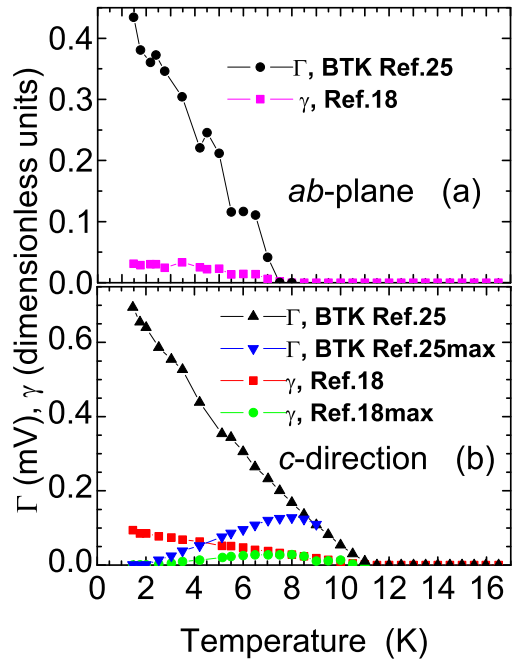}
\caption[]{Temperature dependences of the broadening and pair
breaking parameters for the LuNi$_{2}$B$_{2}$C-Ag point contacts
in the $ab$ plane (a), and along the $c$ axis (b). The index
``max'' marks the ``wings'' fitting (see text). "Wings" fit for
{\it ab}-plane gives $\gamma$=0.} \label{fig7}
\end{figure}

Some experimental curves have a kink marked with an arrow in
Fig.\,\ref{fig6}. This shape is typical of a superconductor with
two gaps of close energies. The temperature dependence of the
broadening (pair breaking) parameter also supports the existence
of two gaps. Since Lu is a nonmagnetic material, it is natural to
expect a negligible broadening (pair breaking) parameter for the
Lu-based borocarbide point contacts. However, that is not the
case, as is evident from the temperature dependences
(Fig.\,\ref{fig7}) of the broadening (pair breaking) parameters
obtained within the two models (Ref.\,\onlinecite{Beloborodko} and
BTK\cite{Plecenik}).

Note that $\gamma$ and $\Gamma$ are different quantities. In the
generalized BTK model \cite{Plecenik} $\Gamma$ is independent of
the gap and has the gap dimensions; whereas, in the model of
Ref.\,\onlinecite{Beloborodko}, $\gamma$ is described as
$\gamma=1/\tau_{S}\Delta$, and, for $\gamma<$ 1, $\gamma$ is
related to the gap $\Delta_{0}$ and the order parameter $\Delta$
as $\gamma=[1-(\Delta_{0}/\Delta)^{2/3}]^{3/2}$. As the
temperature is increased, both $\Gamma$ and $\gamma$ decrease,
and, at temperatures $>$7.5 K for the ab-plane and $>$11 K for the
c-direction, tend to zero. Meanwhile, the one-gap fit starts to
approximate the experimental curve more and more accurately. In
this case the one gap fit with the broadening (pair breaking)
parameter is a tool of describing a certain average gap (order
parameter).

\begin{figure}
\includegraphics[width=8cm,angle=0]{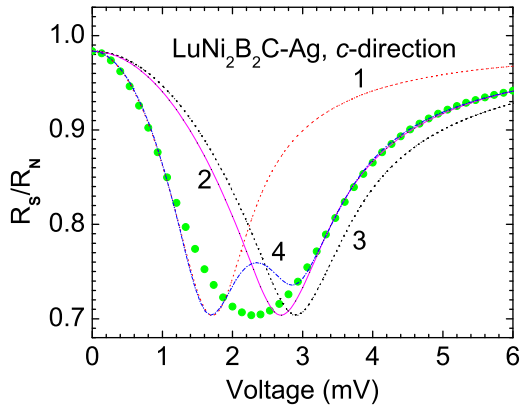}
\caption[]{The estimates of the smallest and largest gaps for the
LuNi$_{2}$B$_{2}$C-Ag point contact in the $c$ direction at
$T=1.47$~K. Curve 1: $\Delta =1.82$~mV ($\Delta _{0}=1.65$~mV),
$\gamma=0.016$, $Z=0.558$; curve 2: $\Delta =2.97$~mV ($\Delta
_{0}=2.693$~mV), $\gamma=0.016$, $Z=0.553$; curve 3: $\Delta
=3.2$~mV ($\Delta _{0}=2.9$~mV), $\gamma=0.016$, $Z=0.558$. The
best two-gap approximation (curve 4, asterisks) is achieved with
the parameters of curves 1 and 3. The relative contribution to the
total conductivity is 60\% from the smaller gap and 40\% from the
larger gap (see the Table). The experimental curve is shown with
solid dots.} \label{fig8}
\end{figure}

The largest and smallest gaps can be estimated by fitting
different portions of the experimental curve. In this case, to
match the amplitudes of the fitting and the experimental curves,
we used another method. Namely, the \textit{y}-coordinates at the
central maximum and near the minima are scaled to coincide
(Fig.\,\ref{fig8}). The best agreement in the central maximum
region can give us an estimate of the smallest gap (curve~1),
while the best agreement in the ``wing'' region yields the largest
gap (curve~2). By "wings" we imply the portion of the experimental
curve at biases higher than the energy gap double-minimum
structure in the dV/dI characteristic, namely at $eV\gtrsim3$. The
estimates however are rather rough.

Further, we follow the method of approximation of $I-V$
characteristics, which was calculated for the $S-I-N$ contacts of
MgB$_{2}$ within a two-band model in Ref.\,\onlinecite{Brinkman}.
There, the total conductivity of the tunnel contact is a sum of
the $\pi-$ and $\sigma-$band conductivities  analyzed by applying
the BTK model. To describe the resulting curve, we also used a
model of two independent parallel-connected point contacts with
different gaps whose conductivities are additive. The
contributions of these conductivities account for the part of the
Fermi surface containing a particular gap. Thus, for the two-gap
model an experimental curve is fitted by the following expression:

\begin{equation}
\frac{dV}{dI}=\frac{S}{\frac{dI}{dV}\left( \Delta _{1},\gamma
_{1},Z\right) K+\frac{dI}{dV}\left( \Delta _{2},\gamma
_{2},Z\right) \left( 1-K\right) } \label{two-gap}
\end{equation}
with a proper choice of the coefficient {\it B}. Here, the
coefficient {\it K} reflects the contribution of the part of the
Fermi surface having the gap $\Delta _1$, {\it S} is the scaling
factor discussed for the one-gap approximation. To obtain the
best agreement with the experiment, the parameters used in this
expression are allowed to differ from those found for fitting of
the separate portions of the experimental curve. It appears that
when the contribution of the smaller gap prevails, the "wings" fit
gives smaller values. In this case, the best result is achieved
with the two-gap fit (Fig.\,8, curve~4, asterisks) using the
parameters for curves~1 and 3, the latter is marked with dots.
Correspondingly, if the larger gap prevails, the central-maximum
fit gives higher values. In Fig.\,6 an example of the two-gap fit
(asterisks) for the {\it ab}-plane is shown.

Although the two-gap fit shows much better agreement in the
regions both of the central maximum and the ``wings'' (Figs. 6, 8
asterisks), it cannot provide a complete description of the
experimental curves, especially at the double-minima structure of
$dV/dI$, and in the $c$ direction (Fig.\,\ref{fig8}). Most likely,
this is because a continuous distribution of the gaps.

The previous estimation of point-contact parameters in the $ab$
plane and in the $c$ direction was made within one-gap and two-gap
models using the equations of Ref.\,\onlinecite{Beloborodko} and
BTK\cite{Plecenik}. The results are presented in the Table.
$\Delta_{1}$, $\Delta_{2}$ are the highest and lowest order
parameters (the same as BTK gaps \cite{Plecenik}) from the two-gap
fit; $\Delta_{01}$ and $\Delta_{02}$ are the energy gaps
corresponding to $\Delta_{1}$, $\Delta_{2}$; $\gamma_{1}$,
$\gamma_{2}$ are the pair breaking parameters of the model
\cite{Beloborodko} for the largest and smallest gaps. (The same
holds for the widening parameters $\Gamma_{1}$, $\Gamma_{2}$ in
the BTK model.\cite{Plecenik}) Independently determined $Z$'s are
the tunneling parameters, which are the same for the largest and
smallest gaps and are thus quite selfconsistent. In the Table the
order parameters (gaps) based on the two-gap fit appear along with
the contributions to the total conductivity from the Fermi surface
region with the corresponding gap. For example, for the $c$
direction contact, the contribution to the conductivity is 40\%
from the region with an order parameter of 3.2 mV and 60\% from
the region with 1.82 mV.

It is seen that within both the models,\cite{Beloborodko,Plecenik}
the dominant contribution to the total conductivity in the $c$
direction is made by the Fermi surface region where the order
parameter is lower. The higher order parameter makes the dominant
contribution in the $ab$ plane. On the average this correlates
with gaps in $ab$ and $c$ directions predicted by the $(s+g)$
model.\cite{review}

Although the part-by-part fitting is rather rough, we tried this
procedure for tracing the temperature dependences of the smallest
and largest gaps. However, more or less definite values were
obtained only at the lowest temperatures because the smallest gap
is estimated within a relatively small part of the experimental
curve (near zero bias). As for the largest gap, its temperature
dependence was traced up to the moment when the calculations over
the entire curve and over the ``wings'' start to give similar
results.

\begin{figure}[htb]
\includegraphics[width=8cm,angle=0]{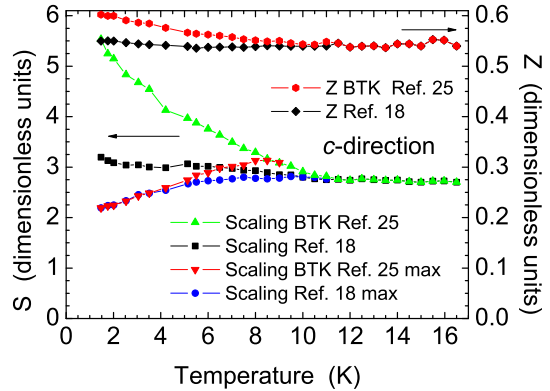}
\caption[]{Temperature dependences of the tunneling parameter $Z$
and the scale factor $S$ for the LuNi$_{2}$B$_{2}$C-Ag point
contact in the $c$ direction obtained with models from
Refs.\,\onlinecite{Beloborodko} and  \onlinecite{Plecenik}. The
scale factor was found from fitting over the entire curve and over
the ``wings'' (Scaling max)} \label{fig9}
\end{figure}

The temperature dependences of the scale factor $S$ in the $c$
direction are shown in Fig.\,\ref{fig9}. They were found for the
largest gap (``wing'' fitting is marked with ``max'' throughout)
and for the average gap (entire-curve fitting) by the equations of
 Ref.\,\onlinecite{Beloborodko} and BTK.\cite{Plecenik}
The temperature dependences of the tunneling parameter $Z$ are
shown for both models. To avoid crowding in the figure, only the
$Z$-values obtained from the entire-curve fitting are shown (the
``wing'' - fitting results are much the same). Although the
fitting was focused on the coincidence of theoretical and
experimental $dV/dI(V)$- curves and no parameters were specially
fixed, we can state that $Z$ and $S$ vary very little in the
whole temperature interval. At $T=9$ K the ``wing'' and
entire-curve fits start to yield similar results.

\begin{figure}[t]
\includegraphics[width=8cm,angle=0]{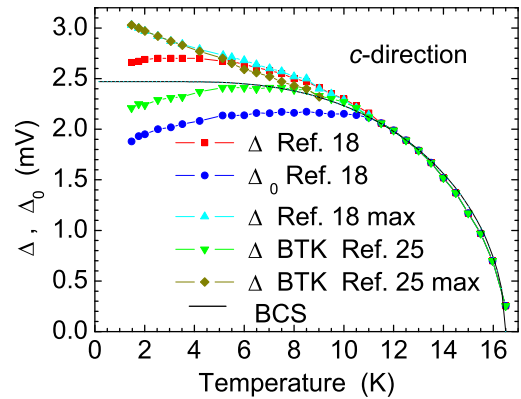}
\caption[]{Temperature dependence of the order parameter $\Delta$
and energy gap $\Delta_{0}$ \cite{Beloborodko} and the BTK energy
gap \cite{Plecenik} for the LuNi$_{2}$B$_{2}$C-Ag point contact in
the $c$ direction. The T-dependence of the largest gap (order
parameter) was estimated from fitting over the entire curve and
over the "wings".} \label{fig10}
\end{figure}

The temperature dependences of the order parameter
\cite{Beloborodko} and the gap in the $c$ direction, obtained by
the entire-curve and the ``wing'' BTK \cite{Plecenik} fittings,
are shown in Fig.\,\ref{fig10} along with the BCS curve. As in the
case of the scaling factor S (Fig.\,\ref{fig9}), deviation starts
at $T\lesssim11$~K, while at higher temperature both fitting
procedures result in the same BCS-like dependence.

\begin{figure}[t]
\includegraphics[width=8cm,angle=0]{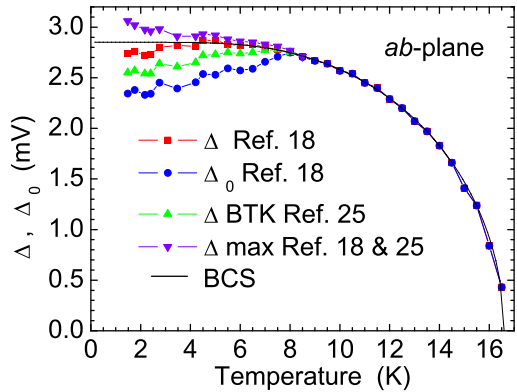}
\caption[]{Temperature dependence of the order parameter $\Delta$
and energy gap $\Delta_{0}$ (calculation by
Ref.\,\onlinecite{Beloborodko}) and the BTK energy gap
(calculation by Ref.\,\onlinecite{Plecenik}) for the
LuNi$_{2}$B$_{2}$C-Ag point contact in the $ab$ plane. The
$T$-dependence of the largest gap (order parameter) was estimated
from fitting over the entire curve and over the ``wings''. The
scatter of the points at $T=1.5-4.2$ K is caused by the contact
instability. The contact resistance becomes stable above 4.5 K.}
\label{fig11}
\end{figure}

The temperature dependences of the order
parameter\cite{Beloborodko} and the
gap,\cite{Beloborodko,Plecenik} for the $ab$ plane contact are
shown in Fig.\,\ref{fig11}. The calculation was made both over the
entire curve and the ``wings''. It is seen that the curves
coincide in a wider temperature interval and follow the
BCS-dependence down to about 8 K. Thus, the gaps in the $ab$ plane
and $c$ direction differ not only in magnitude but in temperature
dependence as well.

Let us consider in more detail the recent STM investigation of
the superconducting gap anisotropy in Lu and Y compounds by means
of STM.\cite{Martinez2} The tunneling spectra obtained in this
work appeared to be impossible to fit in terms of the traditional
BCS theory. For the curve shown in Fig.\,2 of the cited paper,
the zero value of the differential conductivity is seen up to 0.8
mV and the maximum of dI/dV, which corresponds approximately to
the energy gap, is located at 2.3 mV. At the same time, the shape
of the curve is permanent along the surface area much greater
than the coherence length. In Fig.\,2\,b, at energies slightly
less than 2 mV, there is a shoulder, which could be ascribed to a
two-gap spectrum. Unfortunately, this feature is not discussed in
the cited paper. The authors notice that the use of the model,
where the broadening of the BCS density of states is caused by the
finite life time of quasiparticles, implies a non-zero density of
states (or differential conductivity) at infinitesimal bias which
is at odd with the experimental result. Because of that, the
authors of Ref.\,\onlinecite{Martinez2} use a modified density of
state modeled by the energy gap with the Gauss distribution
centered at $\Delta$ with the width $\epsilon$. By investigation
of different parts of the crystal surface, the authors of the
cited paper observe the correlation between the supposed
anisotropy of the gap expressed in $\epsilon$/$\Delta$ units and
the local transition (critical) temperature at the given spot on
the crystal surface. With increase of the critical temperature
under the contact, both the absolute value of the gap and its
anisotropy increase. Such a behavior corresponds to the s-wave
pairing, since in terms of that model the elastic scattering
leads to a decrease of $T_c$ and isotropy of the gap.

In our experiment the critical temperature coincides for both
directions with the bulk value, as one could expect in a crystal
with undisturbed surface. Our parameter $\Gamma$ allows one to
estimate quantitatively the gap anisotropy and is similar to the
parameter $\epsilon$/$\Delta$ in Ref.\,\onlinecite{Martinez2}.
Namely, $\Gamma$ is about 1.5 times greater in the $c$ direction
compared with that in the $ab$ plane. The temperature interval
where our gap dependences deviate from BCS curve is also about
1.5 times larger in the $c$ direction compared with that in the
$ab$ plane. Hence, we may conclude that the anisotropy of the
superconducting gap in the $c$ direction is noticably greater than
in the $ab$ plane.

Note that neither of the techniques\cite{Beloborodko,Plecenik} can
describe adequately the presumed situation because in both cases
the discrepancy is determined by the distribution of the gap over
the Fermi surface. Taking into account the poor efficiency of the
one-gap fit at low temperatures (Fig.\,5) and the deviation of the
fitting curve at the minima of $dV/dI$ for the two-gap fit (see
Figs.\,6,\,8 which give approximately the same discrepancy factor
F as for the lowest temperature in Fig.\,5), we can assert that
the superconducting gap varies \emph{continuously} over the Fermi
surface. It is therefore most reasonable to describe such curves
in terms of a number of parallel-connected point contacts having
different gaps. Their contribution to the total conductivity can
qualitatively account for the region of the Fermi surface
containing a particular gap, as was considered in
Ref.\,\onlinecite{Martinez}.

\begin{table*}[]
\begin{center}
\caption[]{The fitting results for superconducting parameters
along two mutually perpendicular directions, using
Ref.\,\onlinecite{Beloborodko} and the BTK theory.\cite{Plecenik}
$\Delta_1,\Delta_2$ are the order parameters in the two-gap
fitting; whereas, the subindex 0 corresponds to the energy gaps in
the model of Ref.\,\onlinecite{Beloborodko}. The pair breaking
parameters $\gamma_{1,2}$ and the broadening parameter
$\Gamma_{1,2}$ correspond to Refs.\,\onlinecite{Beloborodko} and
\onlinecite{Plecenik}, respectively. $Z$ is the tunneling
parameter. The temperature corresponds to that of the experiment. The
relative contributions of each superconducting order parameter and
gap are given in percent. $\Delta$ and $\Gamma$ are given in meV,
$\gamma$ and  $Z$ are dimensionless.}
\begin{tabular}{p{38pt}p{31pt}p{32pt}p{31pt}p{32pt}p{31pt}p{32pt}p{25pt}p{32pt}p{47pt}p{47pt}p{37pt}p{37pt}}
\hline\hline \raisebox{-3.00ex}{}& \multicolumn{8}{c}{\bf Two-gap
fit} &
\multicolumn{4}{c}{\bf One-gap fit}  \\
& \multicolumn{4}{l}{Ref.\,\onlinecite{Beloborodko}} &
\multicolumn{4}{l}{BTK \cite{Plecenik}} &
\multicolumn{2}{l}{Ref.\,\onlinecite{Beloborodko}} &
\multicolumn{2}{l}{BTK \cite{Plecenik}}  \\
& \multicolumn{2}{l}{$c$} & \multicolumn{2}{l}{\textit{ab}} &
\multicolumn{2}{l}{$c$} & \multicolumn{2}{l}{\textit{ab}} & $c$&
\textit{ab}& $c$& \textit{ab} \\
\cline{1-13} $\Delta _{1}$& 3.2& \raisebox{-1.50ex}{40{\%}}& 3.03&
\raisebox{-1.50ex}{60{\%}}& \raisebox{-1.50ex}{2.65}&
\raisebox{-1.50ex}{50{\%}}& \raisebox{-1.50ex}{3}&
\raisebox{-1.50ex}{62{\%}}&     2.66& 2.74&
\raisebox{-1.50ex}{2.25}&
\raisebox{-1.50ex}{2.55} \\
$\Delta _{01}$& 2.9& & 2.92& & & & & & 1.88& 2.34& & \\
$\Delta _{2}$& 1.82& \raisebox{-1.50ex}{60{\%}}& 2&
\raisebox{-1.50ex}{40{\%}}& \raisebox{-1.50ex}{1.7}&
\raisebox{-1.50ex}{50{\%}}& \raisebox{-1.50ex}{2.2}&
\raisebox{-1.50ex}{38\%}&
\multicolumn{4}{}{\raisebox{-1.50ex}{}} \\
$\Delta _{02}$& 1.65& & 1.82& & & & & &
\multicolumn{4}{p{170pt}}{}  \\
$\gamma _{1},\Gamma _{1}$ & \multicolumn{2}{p{64pt}}{0.016} &
\multicolumn{2}{p{64pt}}{0.004} & \multicolumn{2}{p{64pt}}{0.4} &
\multicolumn{2}{p{64pt}}{0.1} &
\raisebox{-1.50ex}[0cm][0cm]{0.094}&
\raisebox{-1.50ex}[0cm][0cm]{0.031}&
\raisebox{-1.50ex}[0cm][0cm]{0.655}&
\raisebox{-1.50ex}[0cm][0cm]{0.434} \\
$\gamma _{2},\Gamma _{2}$& \multicolumn{2}{p{64pt}}{0.016} &
\multicolumn{2}{p{64pt}}{0.012} & \multicolumn{2}{p{64pt}}{0.4} &
\multicolumn{2}{p{64pt}}{0.14} & & & & \\
$Z$& \multicolumn{2}{p{64pt}}{0.558} &
\multicolumn{2}{p{64pt}}{0.765} & \multicolumn{2}{p{64pt}}{0.59} &
\multicolumn{2}{p{64pt}}{0.77} & 0.55& 0.745& 0.6&
0.804 \\
$T$ (K)& \multicolumn{2}{p{64pt}}{1.47} &
\multicolumn{2}{p{64pt}}{1.48} & \multicolumn{2}{p{64pt}}{1.47} &
\multicolumn{2}{p{64pt}}{1.48} & 1.47& 1.48& 1.47&
1.48 \\
\hline\hline
\end{tabular}
\label{tab1}
\end{center}
\end{table*}

Let us illustrate this approach by fitting the experimental curves
for the $c$- and $ab$-directions at the lowest temperatures, where
the discrepancy is  the highest (Figs.\,6,\,8). Just as the
two-gap fit (\ref{two-gap}), the multigap fit is expressed by
\begin{equation}
\frac{dV}{dI}=\frac{S}{\frac{dI}{dV}\left( \Delta _{1},Z\right)
K_{1}+\ldots+\frac{dI}{dV}\left( \Delta _{n},Z\right) K_{n} },
\end{equation}
where $K_{1}+K_{2}+\ldots+K_{n}=1$, $K_{i}$ is proportional to the
part of the Fermi surface with the gap $\Delta_{i}$, and $S$ is
the scaling factor. The broadening parameter ($\gamma$ or
$\Gamma$) was taken to be zero, and the barrier factor $Z$ was the
same for all the contributions. In this approximation the
approaches of Refs.\,\onlinecite{Beloborodko} and
\onlinecite{Plecenik} are the same, and the order parameter
coincides with the energy gap. As a first approximation, it is
supposed that the coefficients $K_{i}$ are located in the curve
which is a superposition of two peaks (Fig.\,12). The maxima of
these peaks are at energies $\Delta_{k}$ and $\Delta_{p}$, and
their slopes are described by asymmetrical Gaussian distributions
with a width at half height of $\sigma_{1},\sigma_{2},\sigma_{3}$
and $\sigma_{4}$. The heights of each peak are $h_{1}$ and
$h_{2}$, respectively. By varying all the fitting parameters
($\Delta_{i},h_{i},\sigma_{k}$ for $i=1,2$ and $k=1-4$) we
minimized the average r.m.s. deviation. This procedure includes
manually offsetting dots of the K($\Delta$) fitting curve in the
intermediate steps of the adjustment.
\begin{figure}[htb]
\includegraphics[width=8cm,angle=0]{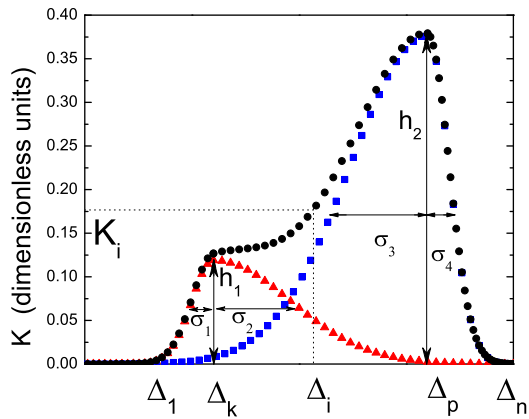}
\caption[]{Modeling of the anisotropic superconducting energy gap
distribution over the Fermi surface with multiple parallel
junctions of different gaps. The contribution of a particular
energy gap correlates with its part over the Fermi surface. The
distribution of a particular contribution to the total
conductivity is described by the $K_{i}(\Delta_{i})$ dependence. The
shape of the distribution $K_{i}(\Delta_{i})$ is simulated by
superposition of two peaks. Each peak is described by the
following parameters: the energy of the central maximum $\Delta_{k}$
or $\Delta_{p}$, half-width of the right and left slopes
$\sigma_{1},\sigma_{2}$ or $\sigma_{3},\sigma_{4}$, and the
heights $h_{1}$ and $h_{2}$, correspondingly. By variation of
these 8 parameters we attempt to reach the best coincidence
between the experimental and theoretical curves. } \label{fig12}
\end{figure}

The result is shown in Figs.\,13 and 14 for the $c$- and
$ab$-directions, respectively. One can see that the multigap fit
with 8 fitting parameters and a continuous distribution of the
gaps matches the experimental curve very well. One should not
trust this fitting literally, but it shows that the width of the
gap distribution can be quite appreciable, down to small values,
which agrees with the $(s+g)$ model of the gap nodes in some
directions.\cite{review}
\begin{figure}[htb]
\includegraphics[width=8cm,angle=0]{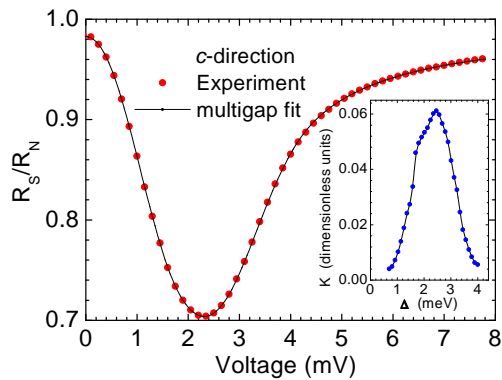}
\caption[]{Approximation of the experimental curve by the use of
the multigap model for $c$ direction. $T=1.47$ K, $Z=0.559$.
Inset: the contribution to the total conductance from different
parts of the Fermi surface with different gaps. The distribution
is modeled by 31 equidistant points, $\Delta_{min}$=0.7 meV,
$\Delta_{max}$=4~meV. The resulting curve exibits two peaks with
maxima at 1.8-2~meV and 2.5 meV. } \label{fig13}
\end{figure}
\begin{figure}[htb]
\includegraphics[width=8cm,angle=0]{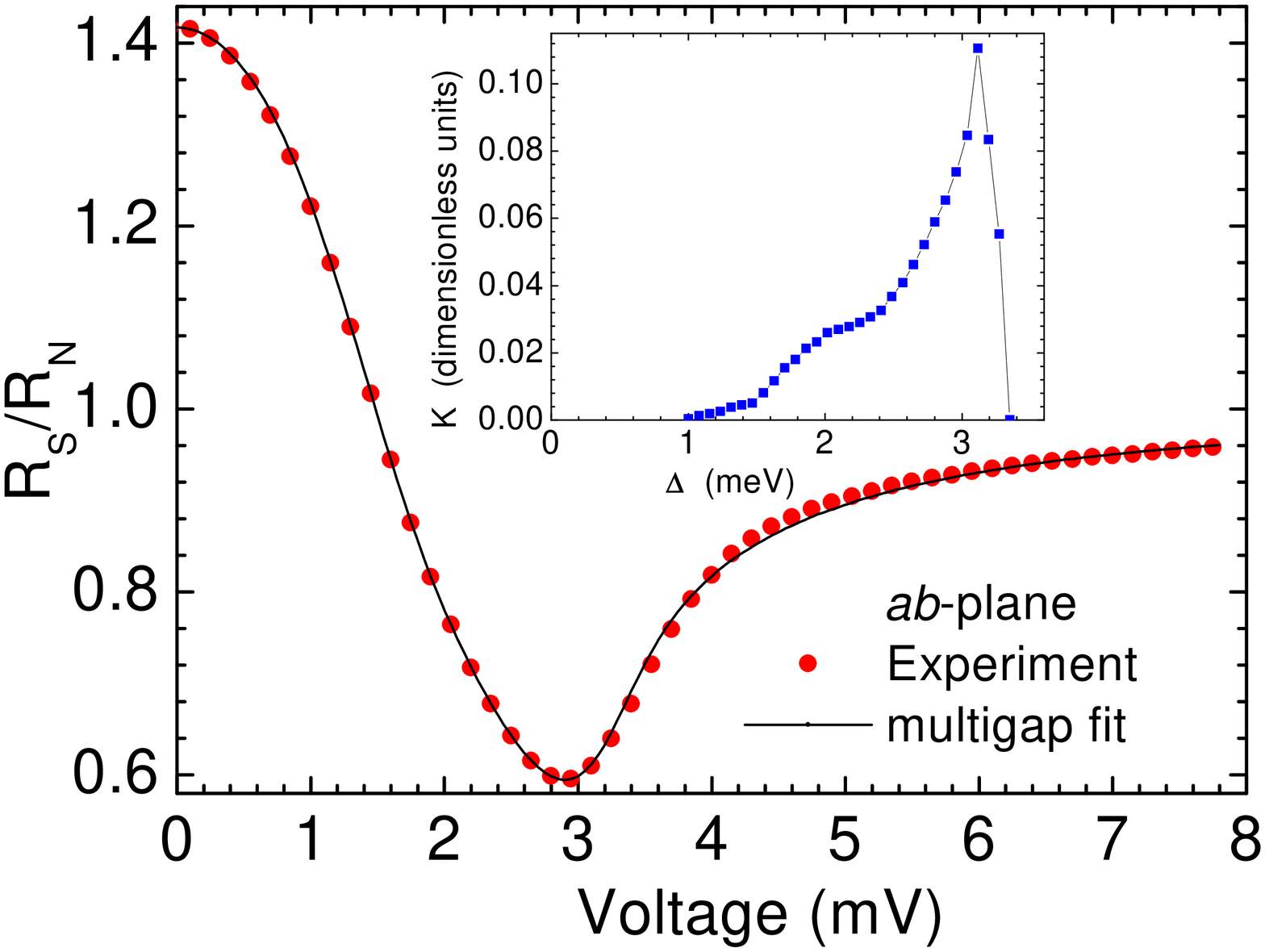}
\caption[]{Approximation of the experimental curve by the use of
the multigap model for the $ab$ plane. $T=1.48$ K, $Z=0.755$.
Inset: the contribution to the total conductance from different
parts of Fermi surface with different gaps. The distribution is
modeled by 31 equidistant points, $\Delta_{min}$=1.0 meV,
$\Delta_{max}$=3.35 meV. The resulting curve exibits two peaks
with maxima at 2 meV and 3.1 meV. } \label{fig14}
\end{figure}

\section{Conclusion}

In this study the anisotropy of the superconducting energy gap was
measured for the first time on LuNi$_{2}$B$_{2}$C in the $ab$
plane and $c$ directions. It is found that at low temperatures the
experimental curves should be described assuming a gap
distribution over the Fermi surface. Within the two-gap model, the
largest contribution to the total conductivity is made by the
Fermi region with a smaller gap in the $c$ direction and by the
Fermi region with a greater gap in the $ab$ plane. At
$T\simeq1.5$~K the largest and the smallest order parameters in
the $c$ direction are $\Delta_{max}$=3.2 meV, $\Delta_{min}$=1.82
meV; in the $ab$ plane these are $\Delta_{max}$=3.03 meV,
$\Delta_{min}$=2.0 meV. An attempt to fit the low temperature
experimental curves shows that the gap is distributed starting
from 0.8 meV. This corresponds well to the recent STM observations
in Ref.\,\onlinecite{Martinez2} discussed above. We have found the
deviation of the temperature dependent gap from the BCS theory for
both tested direction. This deviation is connected with the
impossibility to describe the gap distribution in terms of the
one-gap model, at least at low temperatures. The temperature
range where the described deviation is observed for the $c$
direction, is about 1.5 times greater (1.5 $\div$ 11.5\,K) than
that in the $ab$ plane (1.5 $\div7.5$\,K). The broadening
parameter $\Gamma$, allowing quantitative estimation of the degree
of anisotropy, is also bigger in the $c$ direction.

The experimental results are described on the basis of the
generalized BTK model \cite{Plecenik} and the Beloborod'ko
theory,\cite{Beloborodko} considering the electrical conductivity
of ballistic $S-c-N$ point contacts in the presence of an
arbitrarily penetratable potential barrier and allowing for the
finite lifetime of Cooper pairs. For superconductors with a
multiband electronic structure, the interband transitions of
Cooper pairs may lead to their finite lifetime. Previously, a
two-band model was suggested for nickel borocarbide
superconductors.\cite{Shulga} The theory,\cite{Beloborodko} which
accounts more accurately for the force of pair breaking, may
explain the difference between the order parameter and the gap.

Our next publication will concern the compound ErNi$_{2}$B$_{2}$C
which exhibits a magnetic transition near T=6\,K. In this context
a detailed analysis and comparison of the two theoretical
approaches \cite{Beloborodko,Plecenik} applied to
LuNi$_{2}$B$_{2}$C are of paramount importance for understanding
the results that can be obtained on magnetic ErNi$_{2}$B$_{2}$C.

\section*{Acknowledgments}
The single crystal samples for this study were graciously provided
by P.C. Canfield and S.L. Bud'ko at Ames Laboratory and Iowa State
University. Much of the work reported here was supported in part
by the Robert A. Welch Foundation $(Grant\,A-0514)$, the
Telecommunications and Informatics Task Force at Texas A\&M
University, the Texas Center for Superconductivity and Advanced
Materials at the University of Houston $(TCSAM)$ and the National
Science Foundation ($Grants\, DMR-0103455$ and $DMR-0111682)$.
Partial support of U.S. Civilian Research and Development
Foundation for the Independent States of the Former Soviet Union
(AGREEMENT No.\,UP1-2566-KH-03) is acknowledged.

\end{document}